\documentclass[preprint,prd,nofootinbib,tightenlines,amsmath]{revtex4}

\usepackage{epsfig}
\usepackage{graphics,color}      
\usepackage{verbatim}
\usepackage{amsmath}    
\catcode`\@=11
\def\sla@#1#2#3#4#5{{%
 \setbox\z@\hbox{$\m@th#4#5$}%
 \setbox\tw@\hbox{$\m@th#4#1$}%
 \dimen4\wd\ifdim\wd\z@<\wd\tw@\tw@\else\z@\fi
 \dimen@\ht\tw@
 \advance\dimen@-\dp\tw@ \advance\dimen@-\ht\z@
 \advance\dimen@\dp\z@
 \divide\dimen@\tw@ \advance\dimen@-#3\ht\tw@
 \advance\dimen@-#3\dp\tw@ \dimen@ii#2\wd\z@
 \raise-\dimen@\hbox to\dimen4{%
 \hss\kern\dimen@ii\box\tw@\kern-\dimen@ii\hss}%
 \llap{\hbox to\dimen4{\hss\box\z@\hss}}}}

\def\slashed#1{%
 \expandafter\ifx\csname sla@\string#1\endcsname\relax
{\mathpalette{\sla@/00}{#1}}
\fi}
\def\declareslashed#1#2#3#4#5{%
 \expandafter\def\csname sla@\string#5\endcsname{%
#1{\mathpalette{\sla@{#2}{#3}{#4}}{#5}}}}
 \catcode`\@=12
\declareslashed{}{/}{.08}{0}{D}
 \declareslashed{}{/}{.1}{0}{A}
 \declareslashed{}{/}{0}{-.05}{k}
 \declareslashed{}{/}{.1}{0}{\partial}
 \declareslashed{}{\not}{-.6}{0}{f}

\def\lsim{\mathrel {\vcenter {\baselineskip 0pt \kern 0pt
    \hbox{$<$} \kern 0pt \hbox{$\sim$} }}}
\def\gsim{\mathrel {\vcenter {\baselineskip 0pt \kern 0pt
    \hbox{$>$} \kern 0pt \hbox{$\sim$} }}}

\begin{document}

\baselineskip=15pt

\preprint{BNL-HET-07/20}

\hspace*{\fill} $\hphantom{-}$

\title{Search for RS gravitons via $W_L W_L$ decays}

\author{Oleg Antipin, David Atwood}
\email{oaanti02@iastate.edu}
\email[]{atwood@iastate.edu}
\affiliation{Department of Physics and Astronomy, Iowa State University, Ames, IA 50011, USA}

\author{Amarjit Soni}
\email{soni@bnl.gov}
\affiliation{Brookhaven National Laboratory, Upton, NY 11973, USA\\}

\date{\today}

\begin{abstract}                       

The original Randall-Sundrum (RS) model with a warped extra dimension along with extensions provides the possibility for a simultaneous solution to Planck-weak hierarchy problem as well as the flavor puzzle in the Standard Model (SM). The most distinctive feature of this scenario is the existence of Kaluza-Klein (KK) gravitons whose masses and couplings to the SM fields are set by the TeV scale. 
In some realistic versions of this framework, the largest coupling of the gravitons to the observed particles is to the top quark and unphysical Higgses ($W^{\pm}_L$ and $Z_L$) with the KK graviton (G) masses predicted to be $\gsim$ 4 TeV. We extend earlier works on the KK graviton decays to 
the $t\bar{t}$ final state and to the ``gold-plated'' $Z_L Z_L$ 
modes (with each Z decaying to $e^+ e^-$ or to $\mu^+ \mu^-$)
by studying the resonant production of the gravitons and their subsequent decay to $W_L W_L$ pair. 
We find that with 300 $fb^{-1}$ integrated luminosity of data the semileptonic  $G \to W (\to l \nu_l) W(\to 2 jets)$ mode offers a 
good opportunity to search for the RS KK graviton mode 
with mass lighter than $\sim$ 3-3.5 TeV at the CERN LHC. 
Efficient WW mass reconstruction in the semileptonic mode combined with an analysis of dilepton mass distribution in the purely leptonic channel, $pp \to W (\to l \nu_l) W(\to l^\prime \nu_{l^\prime})$  may help to observe KK $Z^\prime$ and KK graviton separately.
Suitably defined average energy of the charged lepton in the semileptonic mode may be used to 
distinguish decays from longitudinal versus transverse W-bosons.

\end{abstract}

\pacs{PACS numbers: }

\maketitle

\section{Introduction}

String theories inspired particle physicists to solve the problems present in the SM by introducing extra dimensions. Arkani-Hamed, Dimopolulos and Dvali (ADD) \cite{ArkaniHamed:1998rs} proposed existence of $n$ large extra-dimensions with factorizable geometry. In their construction SM fields are confined to our four-dimensional world. 
In this paper we focus instead 
on the Randall-Sundrum framework which uses the idea of a warped extra dimension \cite{Randall:1999ee}. Both scenarios predict existence of KK gravitons. Coupling of each individual ADD graviton to the SM field is suppressed by the Planck scale, but summation over almost continuous spectrum of them compensates for the suppression. RS gravitons, in contrast, have their masses and couplings at the TeV scale and therefore should appear in experiment as widely separated resonances.

The original RS model as well as all of its extensions are based on 
a slice 
of Ad$S_5$ space. At the endpoints of this five-dimensional space two branes are placed which are usually labeled as an ultraviolet (UV) brane and a Planck brane and the large hierarchy of scales is solved by a geometrical exponential factor. Postulating modest-sized $5^{th}$ dimension with radius R and curvature k the TeV/Planck $\sim e^{-k\pi R}$  ratio of scales can be numerically obtained by setting kR $\approx$11. If we assume that all SM fields are localized on the TeV brane, higher dimensional operators in the 5D effective theory will give large contributions to flavor changing neutral current processes and electroweak precision observables. 
Also to solve 
hierarchy in the observed fermion masses 
we need to have different fermion Yukawa couplings for different flavors. 
A natural way to accomplish this proposed by \cite{Davoudiasl:1999tf,Pomarol:1999ad} is to allow SM fields to propagate in the extra dimension. In this scenario light fermions are localized near the Planck brane while heavier ones near the TeV brane and, thus, the problem of fermion masses is solved since Higgs field, localized at the TeV brane will couple weakly to former and strongly to latter fermions. As a consequence, the KK 
graviton whose profile is peaked at the TeV brane will couple mostly to the top quark as well 
as to 
the Higgs (or, by equivalence theorem, to the longitudinal W and Z bosons)~\cite{Fitzpatrick:2007qr,Agashe:2007zd,Davoudiasl:1999jd}.

Thus, the promising channels to observe RS gravitons are those where produced gravitons are decaying to fields localized near the TeV brane.
Search for the KK gravitons using its decays to the top quarks was performed in \cite{Fitzpatrick:2007qr}. The 4-lepton signal through the decay to a pair of $Z_L$'s was studied in \cite{Agashe:2007zd}. Reconstruction possibility of the Z's via their leptonic decays makes this a uniquely clean mode. Both analyses concluded that with $\sim$ 100-300$fb^{-1}$ of data provided by LHC the gravitons of masses up to $\sim$2 TeV can be probed. 

In this paper we will study purely leptonic $G\to W_L W_L \to l \bar{\nu}_l l^\prime \nu_{l^\prime}$ and semileptonic modes, i.e. ($W\to l \nu$)($W\to jets$) from the decay of $W_L$ pair. 
Our analysis suggests that we may be able to observe RS KK graviton mode with mass up to about $\sim$ 3-3.5 TeV
as well as to separate its contribution from that of the RS KK $Z^{\prime}$; thus, if they exist at all, providing strong evidence in favor of the RS 
framework~\cite{adps}.
Our strategy relies on the fact that in the class of models we are working $m_1^G\approx 1.5 m_1^{Z^{\prime}}$ for the lightest KK masses of the graviton and the gauge fields \cite{Davoudiasl:2000wi}. Thus, since $Z^{\prime}$ has only 2/3 of the graviton mass and since cross-section falls quickly as we go up in mass of the resonance being produced, we may expect that 
the gauge KK modes would be more 
accessible~\cite{Agashe:2006hk,Lillie:2007yh,Agashe:2007ki}.  
Then, by making use of the above relationship between masses, we may look for the presence of the graviton in some other mode(s) where they could be well separated.
In particular, we will show that using the purely leptonic mode to observe KK $Z^{\prime}$ (where graviton contribution will be hard to see due to its higher mass) we then may use the semileptonic mode with the knowledge of $Z^{\prime}$ mass to pin down the graviton contribution. 
However, the last channel is challenging as it requires to distinguish two collimated jets from highly boosted W boson from one QCD jet. Therefore, in addition to WW irreducible SM background, we include W + 1 jet background in our study for this mode.

The reason for the enhancement of the graviton signal 
in the WW channel compared to the ZZ mode lies in the fact that the branching ratio (BR) to a $W_L$ pair is twice as big as the BR to a $Z_L$ pair. In addition to that,  Br$(W \to hadrons)\approx$ 2/3 and Br$(W \to l \nu)\approx 1/9$  compared to  Br $(Z \to l^+ l^-)\approx 3.3\%\ $, where $l$ indicates 
each type of the lepton, not sum over them~\cite{tau}.
Also, it is worth mentioning that a RS graviton decays to top quark pairs about $\sim 70 \%\ $ of the time compared to $\sim 15\%\ $ for a $W_L$ pair. The important point for the $t\bar{t}$ final state, however, is that the KK gluon couples to the top pair as well and surpasses graviton 
production~\cite{Agashe:2006hk,Lillie:2007yh}. Also, the reconstruction of such energetic tops far away from the $t\bar{t}$ production threshold might be an additional challenge.

The main experimental problem in using the WW final state with subsequent leptonic decays is the presence of one or two neutrinos. In particular, we most probably will not be able to reconstruct WW mass in the leptonic case; although this channel will be a useful discovery channel to reveal the existence of KK gauge bosons.
Then, we will show that the semileptonic mode should be able to see both signals as they will be well separated due to the significant mass differences mentioned before.

To summarize, our channels allow us to probe first RS KK graviton mode with mass below 3-3.5 TeV and, also, to distinguish it from the contributions of RS spin-1 KK gauge bosons. Certainly, the full establishment of the existence of spin-2 graviton from the RS model will need combined analysis of modes discussed in this paper with other decay modes considered before in the literature \cite{Fitzpatrick:2007qr,Agashe:2007zd}. The role of the $Z_LZ_L$ 
mode is extremely important as this mode is 
forbidden for $Z^{\prime}$'s to 
decay into. Note, though, that 
the ``gold-plated'' nature of this special mode, 
with each Z decaying to $e^+ e^-$ and $\mu^+ \mu^-$,
comes at the price of needing a 
higher luminosity~\cite{adps}. 
Thus, with the strategy discussed above and better statistics in (semi)leptonic modes of W's, we may optimistically have evidence for the RS gravitons.

\section {Model}

We closely follow the model discussed in \cite{Agashe:2007zd} and briefly 
review it here. As discussed above, we allow SM fields to propagate in the extra dimension and distribute fermions along it to generate observed mass spectrum without introducing additional hierarchies. SM particles are identified with zero-modes of 5D fields, and the profile of the fermion in the extra dimension depends on its 5D mass. As was shown before \cite{Grossman:1999ra,Huber:2000ie,Gherghetta:2000qt}, all fermion 5D masses 
are $O$(1) parameters with the biggest one,
among the SM quarks, being that of the top quark. 
To specify the model even further, the top quark is localized near the TeV brane and the right-handed isospin is gauged \cite{Agashe:2003zs}. We consider $t_R$ being on the TeV brane (see discussion of the other possibilities in \cite{Agashe:2007zd}, for example). At the end of the day, we are left with three parameters to be measured experimentally. We define them as $c\equiv k/M_{Pl}$ the ratio of the AdS curvature $k$ 
to the 
Planck mass; $\mu\equiv ke^{-\pi kR}$ 
which monitors gauge KK masses with the first few being (2.45, 5.57,
8.7...)$\times \mu$;
and finally, parameter $\nu\equiv m/k$, which defines where the lightest fermion with bulk mass $m$ is localized. For the $t_R$ on the TeV brane, $\nu_{t_R}\approx0.5$; and the parameters $c$ and $\mu$ will remain free in our analysis.

\subsection{Low energy constraints on model parameters}

We now briefly review constraints placed on the warped extra-dimension 
model with custodial isospin symmetry \cite{Agashe:2003zs}, 
which we adopt in this paper. As it was shown in \cite{Agashe:2003zs},
KK mass scale as low as $\sim$ 3 TeV is allowed by precision electroweak 
data. 
Regions of parameter space that successfully reproduce the fit
to electroweak precision observables with KK excitations
as light as $\sim$ 3 TeV were also studied in \cite{Carena:2006bn}.
Implications of the observed  
B-mixings were discussed in \cite{Chang:2006si}. 
In the model of \cite{Agashe:2003zs} B-mixing is mainly 
accommodated by tree level exchange of KK gluons. In \cite{Chang:2006si}, 
the CP-violating effects on the $B_d$ system were shown to 
provide  $M^{(1)}_{gluon}>$3.7 TeV  constraint 
at 68\% CL.

Phenomenological constraints from lepton-flavor-violations were
discussed in \cite{Agashe:2006iy, Chang:2007uz}. In
\cite{Agashe:2006iy}, ``anarchic'' Randall-Sundrum model of flavor was
studied, and the minimal allowed KK scale of $\sim$ 3 TeV was found to
be permitted for a few points in the natural RS parameter space; but
models with custodial isospin can relax these constraints. 
In
\cite{Chang:2007uz}, after extensive analysis of $B \to K^* l^+ l^{'-}$
modes, only the $B\to K^* ee$ decay was found to have sizable new physics
effects. With negligible SM contributions, current experimental bounds
were translated into the lepton bulk mass parameters. For the first KK
gauge boson mass of 2-4 TeV, 10-20$ \%\ $ deviation from the SM results
were found.
Top quark flavor violations
and B-factory signals were also studied 
in \cite{Agashe:2004ay,Burdman:2003nt,Agashe:2006wa}. 
Finally, in addition to the above mentioned constraints, 
since these frameworks contain beyond the SM 
operators with $(V - A) \otimes (V + A)$ structure,
they generate enhanced contributions 
to $\Delta S=2$ processes \cite{Beall:1981ze}.  
Without further flavor structure these contributions are expected to place a 
lower bound 
on the KK gluon 
mass of $O$(8 TeV)~\cite{Bona:2007vi,Agashe:2007ki}.    


Actually, model building based on the underlying RS ideas
continues to flourish, specially designed to find ways to 
lower the allowed KK-masses in face of the various experimental
constraints.    
In fact, even a somewhat surprising claim
that KK masses as low as 1 TeV, consistent with all
current experimental constraints, may be found 
in \cite{Moreau:2006np}. An interesting variant of the
warped extra dimension based on 5D minimal flavor violation was
recently presented in \cite{Fitzpatrick:2007sa}. The model allows to
eliminate current RS flavor and CP problem with a KK scale as low as 2
TeV. Finally, a volume-truncated version of the RS scenario called "Little Randall-Sundrum (LRS)" model 
was constructed in \cite{Davoudiasl:2008hx}. Assuming separate gauge and flavor dynamics, a number of unwanted contributions to precision electroweak, $Z b\bar b$, and flavor observables were shown to be suppressed in the LRS framework, compared with the corresponding RS case.

Bearing all this in mind, current theoretical constructions 
suggest that it would be difficult to have
KK gauge bosons with masses below 3 TeV (which would imply $m_G \gsim$ 4
TeV); if true then, as seen already in 
other studies~\cite{Agashe:2007zd,Agashe:2007ki} 
and will be shown here too, signals at the LHC for the RS idea
would be extremely difficult to find.
However, in light of the above discussion, it also seems fair to say
that these models are still being developed; and, therefore,
it is not inconceivable that explicit construction(s) will be found
which will allow KK masses lower than 3 TeV without running into
conflict with electroweak precision experiments or with flavor physics.
This point of view, in particular,  was also emphasized in \cite{open}. 
Thus in this paper we will take the view, for now, that it is best to 
search for experimental signatures with the  widest latitude. 

\subsection{Couplings of KK gravitons}
After these brief remarks we can write the couplings relevant to our discussions here. Since the graviton couples to the energy-momentum tensor, all couplings have generic form $C_{00n} h_{\mu\nu}T^{\mu\nu}$ (``00n'' signifies that we are considering only coupling of the nth KK graviton to the SM fields which are zero-modes of the 5D fields). Magnitude of the coupling constants depend on the overlap of the particle wavefunctions in the extra-dimension (effects of the running gravitational coupling due to existence of non-Gaussian fixed point were analyzed in \cite{Hewett:2007st,Litim:2007iu}). We present coefficients $C_{00n}$ in Table \ref{table} along with partial decay widths for dominant decay channels for the lightest KK (n=1) graviton which will be the focus of our analysis; see also \cite{Agashe:2007zd}. The $W_L W_L,Z_L Z_L$ and hh decay channels illustrate equivalence theorem once again (which is valid up to $(M_{W,Z}/m_G)^2$ where $m_G$ is the graviton mass).

\begin{table}
\caption{Couplings of the nth level KK graviton to the SM fields. $t_R$ assumed to be localized on the TeV brane. Parameter $m_1^G$ is the mass of n=1 graviton and $x_1^G=3.83$ is the first root of the first order Bessel function. $N_c=3$ is number of QCD colors.}
\label{table}
\begin{center}
\begin{tabular}{|c|c|c|}
\hline
SM fields&$C_{00n}$& Partial decay widths for n=1 graviton\\
\hline
gg(gluons)&$\frac{c}{2\pi kR\mu}$& negligible\\
\hline
$W_L W_L$&2c/$\mu$&$(cx_1^G)^2 m_1^G/480\pi$\\
\hline
$Z_L Z_L$&2c/$\mu$&$(cx_1^G)^2 m_1^G/960\pi$\\
\hline
$t_R \bar{t}_R$&c/$\mu$&$N_c(cx_1^G)^2 m_1^G/320\pi$\\
\hline
h h&2c/$\mu$&$(cx_1^G)^2 m_1^G/960\pi$\\
\hline
\end{tabular}
\end{center}
\end{table}

 The suppression in coupling of the graviton to the gluons follows because gauge boson has a flat wavefunction and thus its couplings to the graviton is suppressed by the volume of the bulk $\pi kR\approx 35$. For the same reason, decay of gravitons to transverse W and Z bosons as well as photons are suppressed by this volume factor. The masses of the KK gravitons are given by $m_n=x_n \mu$ where $x_n$ is n'th zero of the first order Bessel function.
Notice that we do not need $q\bar{q}$G coupling as it is Yukawa-suppressed and graviton production is dominated by gluon fusion.

In this model the total width of the graviton is found to be $\Gamma_G=\frac{13(cx_1^G)^2 m_1^G}{960\pi}$ which is split between 4 dominant decay modes to $W_L W_L,Z_L Z_L,t_R \bar{t}_R$ and hh in the ratio 2:1:9:1. Taking $c\sim 1$, 
the total graviton width is $\sim 6 \%\ $ of its mass and is very close to the corresponding width for RS KK Z$^{\prime}$ in the same model \cite{Agashe:2007ki}.

\section{Production and decay of KK gravitons}

We are now in position to calculate the 
matrix element for the gg$\to G_n \to W_L W_L$. The details 
can be found elsewhere \cite{Park:2001vk,Agashe:2007zd}:

\begin{equation}
M(g^a g^b \to W_L W_L)= \frac{c^2}{\pi kR \mu^2} \cdot \frac{2A_{+-00}\delta_{ab}}{s-(m_n^G)^2+i\Gamma_n^G m_n^G}
\end{equation}
where, $A_{+-00}=A_{-+00}=\frac{1}{2}(\hat{\beta}^2-2)\hat{s}^2 $sin$^2\hat{\theta}$ is the only independent helicity amplitude for the decay to longitudinal W bosons. W boson velocity $\hat{\beta}^2=1-4M_W^2/\hat{s}$ and all hatted variables refer to the parton center of mass frame. We see that the amplitude has sin$^2\hat{\theta}$ behavior characteristic of the $W_L$ pair in the final state. This implies that our signal events will be concentrated in the central rapidity region, and we will exploit this fact later to separate our signal from SM background.

This amplitude gives the parton level cross-section \cite{Agashe:2007zd}:

\begin{equation}
\frac{d\hat{\sigma}(gg\to W_L W_L)}{dcos\hat{\theta}}= \frac{|M|^2 \hat{\beta}}{512\pi \hat{s}},
\end{equation}
and the proton level cross-section is obtained by convolving parton level cross-section with gluon PDF's:

\begin{equation}
\sigma(pp \to WW)= \int dx_1 dx_2 f_g(x_1,Q^2) f_g(x_2,Q^2)\hat{\sigma}(x_1 x_2 s).
\end{equation}

Note that the total cross-sections for the EW boson final states are related by $\sigma(pp \to G \to W_L W_L)=2 \times \sigma(pp\to G \to Z_L Z_L)$. Numerical results for our 2$\to$2 process can be 
found in 
\cite{Agashe:2007zd} (where the $Z_L$ final state was used) which agrees with our current calculation.

\section{Battling SM background}

We now discuss the relevant decay modes of the W bosons.

If both W's decay hadronically, we face huge QCD background; and, therefore, this mode is unlikely to be useful.
Thus, in the rest of this paper, we concentrate on pure leptonic and semileptonic decay modes of the W pair and consider the former first.

\subsection{Pure leptonic mode: $e^\pm \mu^\mp$ final state}

Due to the significant boost of the W's, neutrino's $p_T$ in this mode will be almost back to back; and, therefore, missing energy information will be lost. We require the W's to decay to different lepton flavors since in SM there is no basic $2 \to 2$ partonic process giving two different high transverse momentum lepton flavors in the final state. After this, leading irreducible SM backgrounds for our $l^+ l^{\prime-}\slashed{E}_T$ final state are $W^+W^-\to l^+ l^{\prime-}\slashed{E}_T$ and $Z/\gamma^* \to \tau^+ \tau^- \to l^+ l^{\prime-}\slashed{E}_T$ (where  $l\neq l^\prime$).

With two neutrino's in the final state, we might not reconstruct the resonant 
W boson mass. However, we will show that even looking only at the leptons (at this point by leptons we mean primarily e and $\mu$; see, however, further discussion on $\tau$'s later) may provide enough data to discover the RS graviton, given a suitable set of cuts. We will show that such cuts give S/B $\gsim$ 1 for mass of the first KK graviton mode $m_1^G\lsim$ 3 TeV.

\subsection{Semileptonic mode}

For semileptonic channel we have highly collimated decay products for both 
W's. For the hadronic side, it implies that the two jets from the 
W decay  are likely to appear as one ``fat'' W-jet. This leads us to consider W + 1 jet which will be the leading background for this decay mode compared to irreducible SM WW production and  W + 2 jets (which will be suppressed due to the 3 body phase-space). On the leptonic side, due to small angular separation between missing neutrino and charged lepton, we may estimate longitudinal (L) component of the $\nu's$ momentum as 

\begin{equation}
p_{\nu}^{L}\approx\frac{\slashed{E}_T p_l^{L}}{p_{T_l}}.
\end{equation} 
Using this collinear approximation, the momentum of the leptonic W is reconstructed and, thus, 
 we can calculate the (presumably) resonant invariant mass of the semileptonic system as $M_{WW}^2=(p_{l\nu}+p_{jj})^2$. Notice that we assumed that leptons are coming from the W decay as the reconstructed leptonic W mass will be zero in the collinear approximation.
Also notice that in this approximation, the $M_{WW}$ measurement error for the TeV energy W bosons is $\sim m_W/E_W\sim$ 0.1.
The accuracy of this invariant mass measurement will also depend on how effectively hadronic W side can be reconstructed. We will elaborate on this at the beginning of section {\bf \ref{chapt5} B}.

\section{Acceptance cuts and results \label{chapt5}}

We now present our results as well as specify selection criteria for signal events. We estimated SM background with the aid of the COMPHEP package \cite{comphepre}.  For our graviton signal we used Mathematica program and partially cross-checked them with COMPHEP. For an additional check, we confirmed results of Ref.\cite{Agashe:2007zd} for $\sigma(pp\to G \to Z_L Z_L)$, as was mentioned before. The CTEQ5M PDF's were used throughout (in their Mathematica distribution package \cite{Pumplin:2002vw} as well as intrinsically called by COMPHEP).

\subsection{Pure leptonic mode: $e^\pm \mu^\mp$ final state}

As a starting point, before imposing any cuts, we reproduced results of Ref.\cite{Accomando:2007xc} which finds  $\sigma(pp \to e^+ \nu_e \mu^- \bar{\nu}_{\mu})\approx$ 610 fb and is dominated by WW production. We cross-checked our WW production results with Ref.\cite{Agashe:2007ki} as well.

We impose basic acceptance cuts as

\begin{equation}
|\eta_l|<3, \hspace{5mm} p_{T_l}>50 GeV, \hspace{5mm} \slashed{E}_T>50 GeV,
\label{lepbasic}
\end{equation}
where $\eta_l$ is the pseudorapidity of the charged lepton.

\begin{figure}[htb]
(a)\includegraphics[width=4.6in]{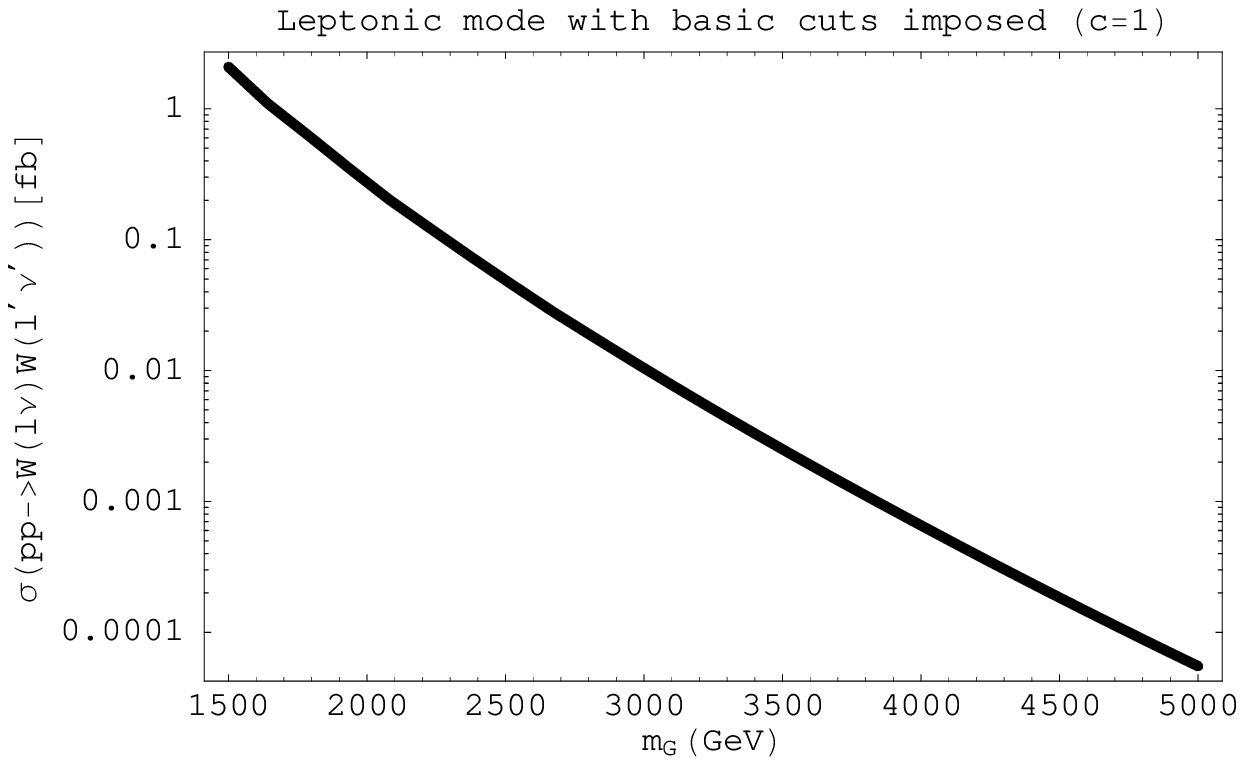} (b)\includegraphics[width=4.6in]{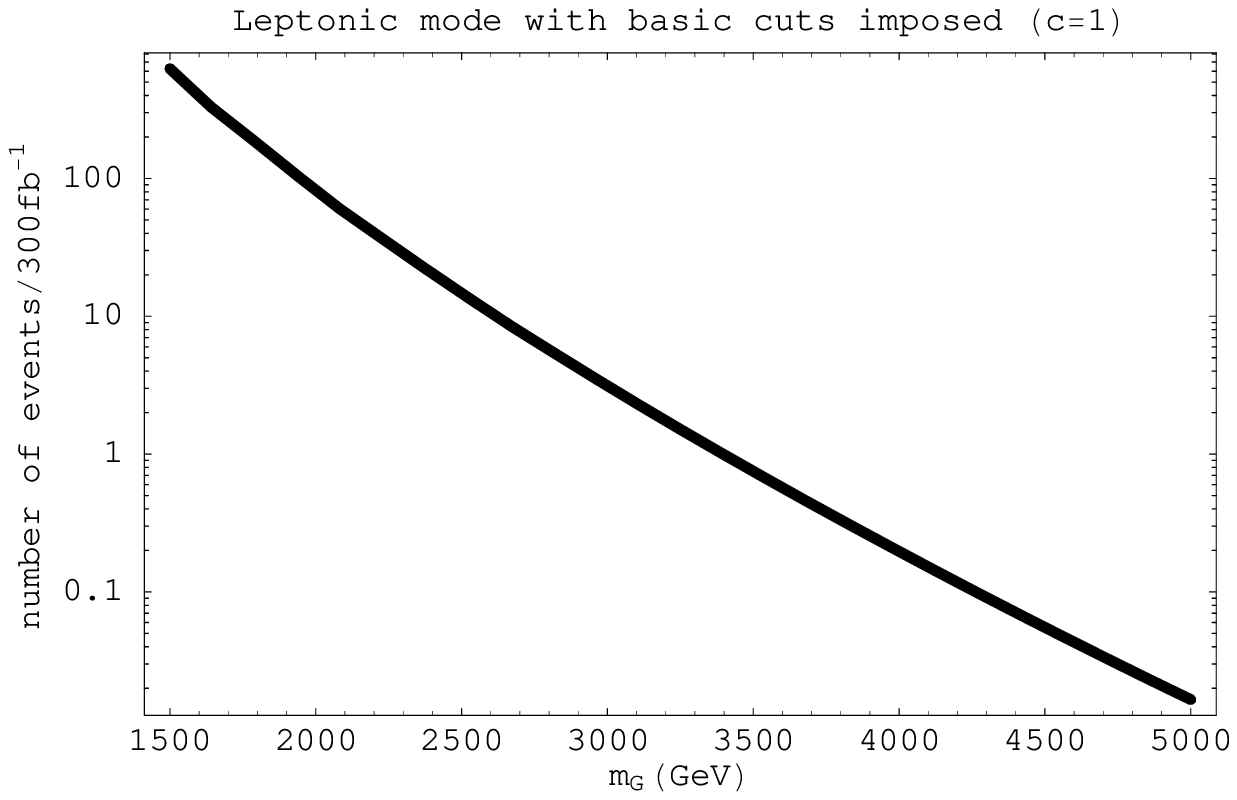}
\centering
\caption{(a) Total signal cross-section for $pp \to l \nu l^{\prime} \bar{\nu}_{l^{\prime}}$, and (b) corresponding number of events for 300 fb$^{-1}$. Basic cuts from Eq.\ref{lepbasic} are applied and c=1. Corresponding SM background is $\approx$ 24 fb and is independent of the graviton mass.}
\label{nocuts}
\end{figure}

In Fig.\ref{nocuts}a,b we show the total cross-section for $pp \to l \nu_l l^\prime \bar{\nu}_{l^\prime}$ (where  $l\neq l^\prime$) and expected number of events per 300 fb$^{-1}$ as a function of $m_G$ for our signal. 
The corresponding SM background is $\approx$ 24 fb and is dominated by 
the WW production with contribution from $Z/\gamma^* \to \tau^+ \tau^- \to l^+ l^{\prime-}\slashed{E}_T$ process being about an order of magnitude smaller, which also is in good agreement with corresponding results of Ref.\cite{Agashe:2007ki}. 

\begin{figure}[htb]
\includegraphics[width=5.0in]{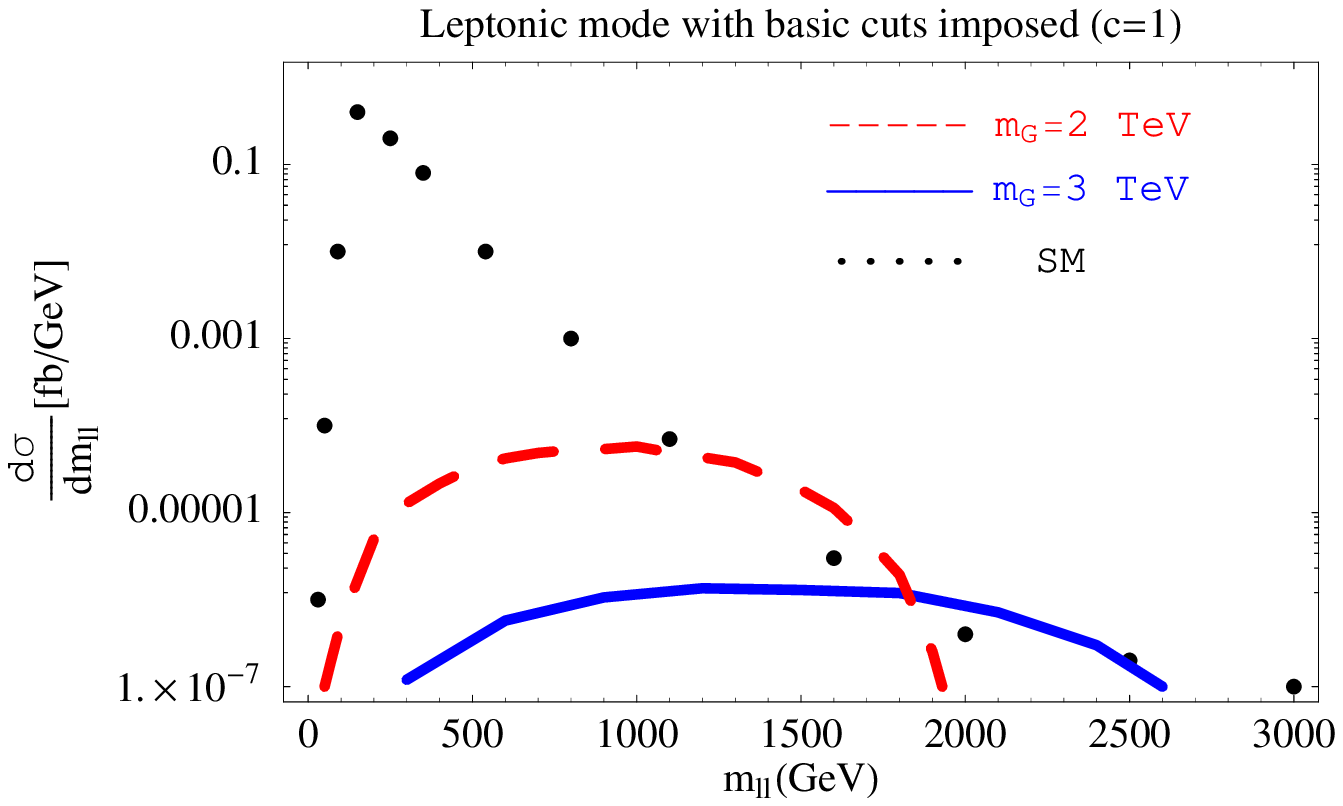} 
\centering
\caption{(Color online) Dilepton invariant mass distributions for graviton masses of 2 TeV (dashed red) and 3 TeV (solid blue). Dotted curve corresponds to the SM background.}
\label{distr}
\end{figure}

We see that as the SM background dominates, we need to look for additional cuts to improve signal observability. Invariant dilepton mass may provide additional information to enhance our S/B ratio. In Fig.\ref{distr} we show dilepton invariant mass distributions for signal and corresponding background where $m_G=$2 TeV and 3 TeV values were chosen. 

We observe that the SM background distributions tend to peak at low dilepton invariant mass while signal events concentrate in the middle mass region dictated by the decay of the very massive object. This allows us to define cuts on dilepton mass. For the masses shown on Fig.\ref{distr}, for example, we have chosen them as
\begin{eqnarray}
m_G=2 \hspace{2mm}TeV:&&  m_{ll^{\prime}}> 1 \hspace{2mm}TeV \nonumber\\
m_G=3 \hspace{2mm}TeV:&&  m_{ll^{\prime}}> 1.5 \hspace{2mm}TeV 
\label{dilcut}
\end{eqnarray}
to improve the statistical significance of the signal further.
Table.\ref{table2} shows the 
statistical results after all the cuts defined above were applied. We notice that the SM background was reduced significantly while the signal was roughly reduced by half. 
Throughout the paper, Poisson statistics CL to observe at least one signal event will be appropriate description if the number of background events is $\lsim$ 10. When needed, these CL are given in brackets next to the corresponding statistical significances in Gaussian statistics. 

In the model we are working, there will also be a contribution to the signal from the KK $Z^{\prime}$. If we use the mass ratio of this model $m_1^G\approx 1.5 m_1^{Z^{\prime}}$, we observe, for example, that a 3 TeV graviton should appear along with a 2 TeV Z$^{\prime}$. Interestingly, we find that the total production cross-section for 2 TeV graviton and $Z^{\prime}$ are very similar in magnitude (it is about 16 fb for $Z^{\prime}$ \cite{Agashe:2007ki} compared to 10 fb for graviton) and shape. Thus, 2 TeV graviton contribution in Fig.\ref{distr} may be numerically viewed as the one coming from Z$^{\prime}$. After this observation, Fig.\ref{distr} represents signal cross-section for 3 TeV graviton along with SM background and 2 TeV RS $Z^{\prime}$. Similarly for 2 TeV graviton, 1.33 TeV $Z^{\prime}$ needs to be considered and so on. As two contributions are mixed up in this channel, it might be easier to ``reserve'' this channel for $Z^{\prime}$, since corresponding graviton contribution will be negligible. Stated differently, if enhancement in dilepton mass due to these states will be observed experimentally, most probably Z$^{\prime}$ will have a dominant effect. Then, Fig.\ref{distr} may be used to define a proper cut on dilepton mass variable to remove this $Z^{\prime}$ background (for example, for 3 TeV graviton $m_{ll}> 2$ TeV will work). Of course, it might happen that the lightest KK $Z^{\prime}$ and graviton masses are actually in different ratio; and we need other measurement(s) to interpret enhancement in dilepton mass. In the next section we will show that semileptonic mode may provide this additional handle.

\begin{table}
\caption{Purely leptonic mode cross-sections [in fb] and S/B ratios after basic and dilepton mass cuts in Eq.\ref{lepbasic} and Eq.\ref{dilcut} were 
imposed.  When the
number of events is low ($\lsim$ 10),
Poisson statistics confidence level is considered a more appropriate
statistical
description and is consequently used.}

\label{table2}
\begin{center}
\begin{tabular}{|c|c|c|c|c|c|}
\hline
2 TeV&Basic cuts& Dilepton mass cut&$\#$ of events/300 fb$^{-1}$&S/B&S/$\sqrt{B}$\\
\hline
Signal&0.22&0.1&30&2.5&8.7\\
Background&24&0.04&12&&\\
\hline
3 TeV&Basic cuts& Dilepton mass cut&$\#$ of events/300 fb$^{-1}$&S/B&S/$\sqrt{B}$\hspace{2mm}(CL)\\
\hline
Signal&0.0087&0.004&1.2&0.6&0.8\hspace{2mm}(64$ \%\ $)\\
Background&24&0.007&2.1&&\\
\hline
\end{tabular}
\end{center}
\end{table}

\subsection{Semileptonic decay}

\begin{figure}[htb]
\includegraphics[width=5in]{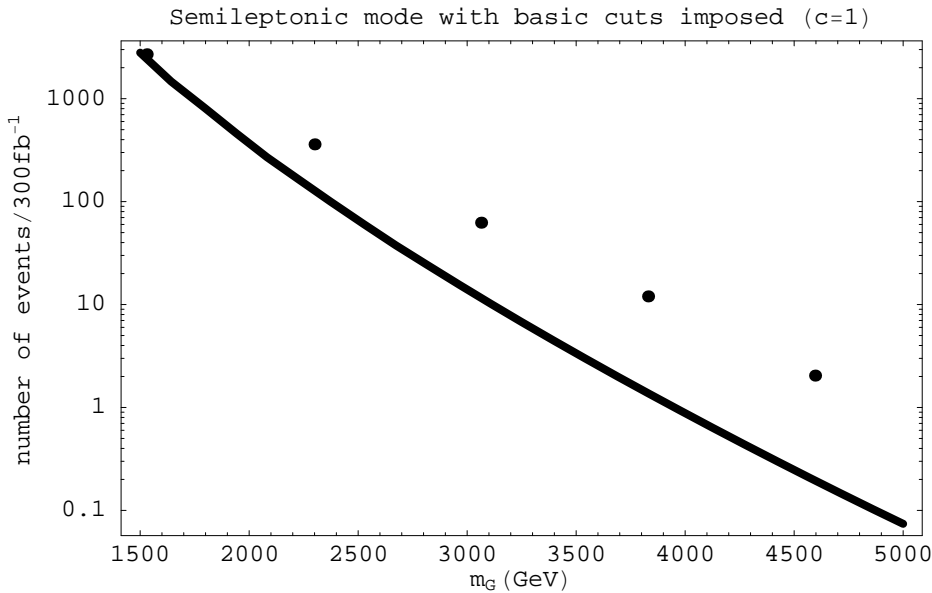}
\centering
\caption{The total signal (solid) and SM background W + 1 jet (dotted) cross-section (integrated in $m_G \pm \Gamma_G/2$ window) after cuts specified in Eq.\ref{slcuts} were applied.}
\label{WW1}
\end{figure}

As discussed above, leptonic W momentum for this mode can be reconstructed; and W + 1 jet is a leading background. As in the case of the leptonic mode, we define basic selection cuts as

\begin{equation}
|\eta_{l,j}|<1, \hspace{5mm} p_{T_l}>50 GeV, \hspace{5mm} \slashed{E}_T>50 GeV, \hspace{5mm} p_{T_j}>100 GeV ;
\label{slcuts}
\end{equation}
and in Fig.\ref{WW1} we show the expected number of signal and background events, both integrated over one half of the graviton width. Assuming again that $m_G \approx 1.5 m_{Z^{\prime}}$, the Z$^{\prime}$ contribution is negligible in this WW invariant mass window because Z$^{\prime}$ and graviton total widths are $\sim$ 5$\%\ $ of their mass, while the mass difference between Z$^{\prime}$ and graviton is $\sim$ 50$\%\ $ of Z$^{\prime}$ mass; this also assumes that the $Z^{\prime}$ mass is established by the pure leptonic mode.

We see that the background is severe; and, therefore, its reduction is a serious and challenging issue. One quantity that may help to resolve the problem is a jet-mass, which is the combined mass of the vector sum of 4-momenta of all hadrons making up the jet. For the signal, we expect jet-mass to peak at $M_W$. Along these lines, as it was shown in Ref.\cite{Agashe:2007ki}, the cut on the jet-mass 75 $< M_{jet}<$ 125 GeV gives a substantial rejection of the background events ($\approx 70 \%\ $) while accepting most of the signal. Also, EM calorimeter, due to its finer segmentation, may allow to improve jet-mass resolution, since signal W events are expected to have two separated EM cores. For further discussion on this issue, we refer to Ref.\cite{Agashe:2007ki,Colas:2005jn,Benchekroun:2001je,Skiba:2007fw}.

\begin{figure}[htb]
(a)\includegraphics[width=4.6in]{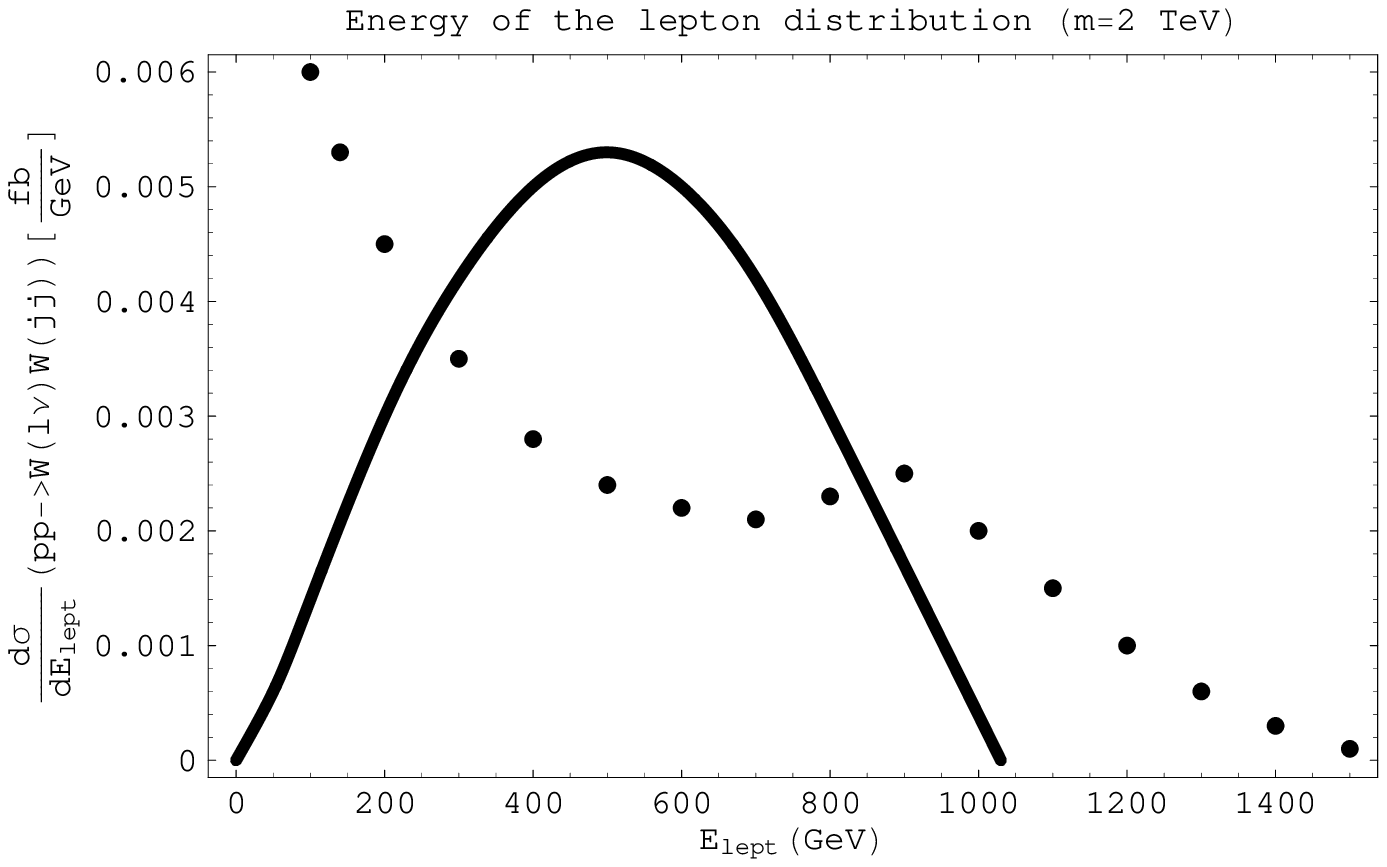} (b)\includegraphics[width=4.6in]{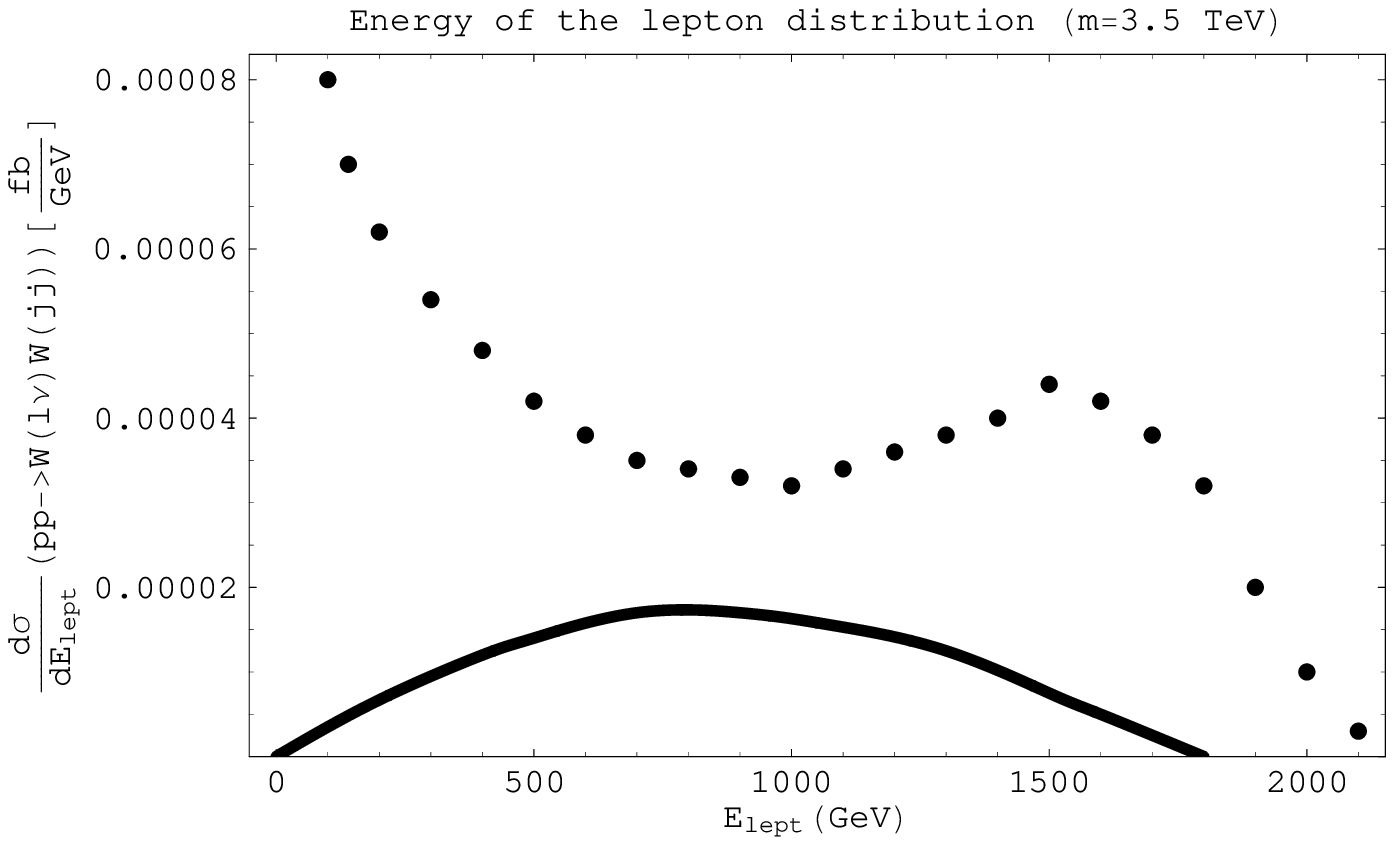}
\centering
\caption{ Differential lepton energy distribution for the signal (solid) and SM  W + 1 jet background (dotted) (integrated in $m_G \pm \Gamma_G/2$ window) after cuts specified in Eq.\ref{slcuts} were applied for (a) $m_G$=2 TeV and  (b) $m_G$=3.5 TeV.}
\label{WW2}
\end{figure}

In an attempt to specify the selection cuts further, in Fig.\ref{WW2} (note that the scale is linear) we show the lepton energy distribution for graviton masses $m_G=2$ TeV and $m_G$=3.5 TeV  for the same conditions as in Fig.\ref{WW1}. We observe that by defining appropriate cuts on lepton energy signal observability can be improved. We define them as

\begin{eqnarray}
m_G=2 \hspace{2mm}TeV:&& 0.2 \hspace{2mm} TeV < E_{lepton}< 1 \hspace{2mm}TeV \nonumber\\
m_G=3.5 \hspace{2mm}TeV:&& 0.5 \hspace{2mm} TeV < E_{lepton}< 1.4 \hspace{2mm}TeV 
\label{lepen}
\end{eqnarray}
and show resulting statistics in Table.\ref{table3}. We observe that with 300fb$^{-1}$, it is possible to reach 1$\sigma$ effect for 3.5 TeV graviton.

\begin{figure}[htb]
(a)\includegraphics[width=4.6in]{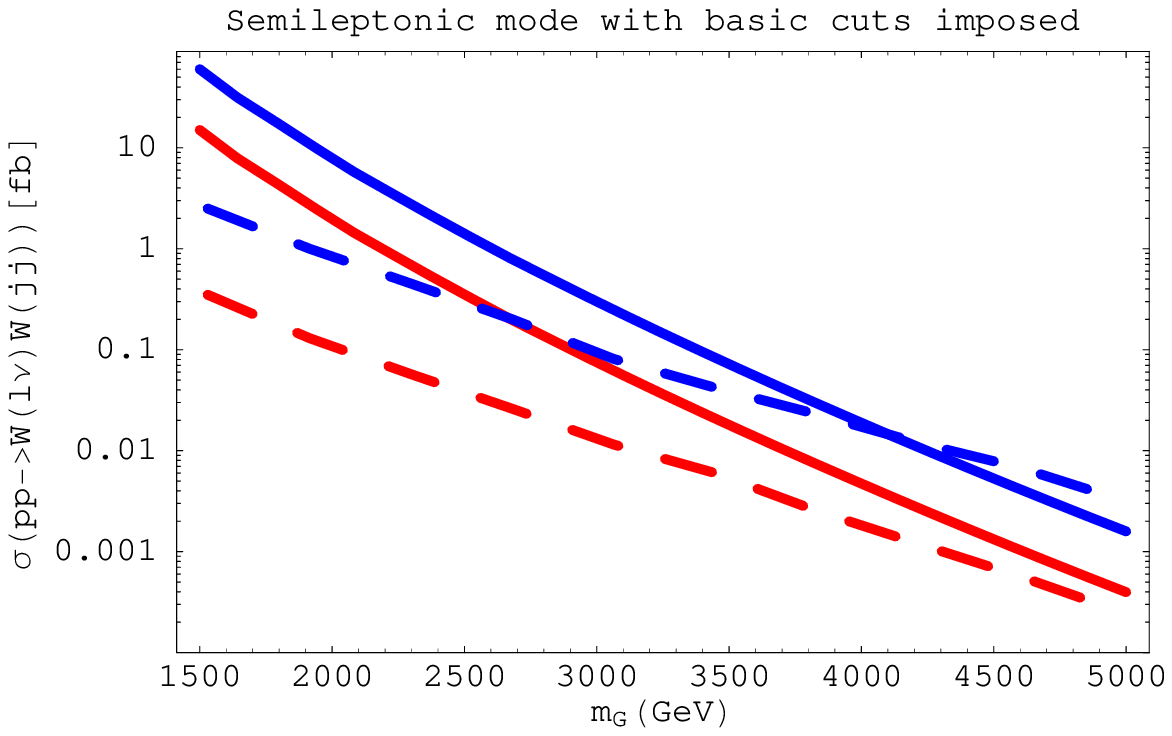} (b)\includegraphics[width=4.6in]{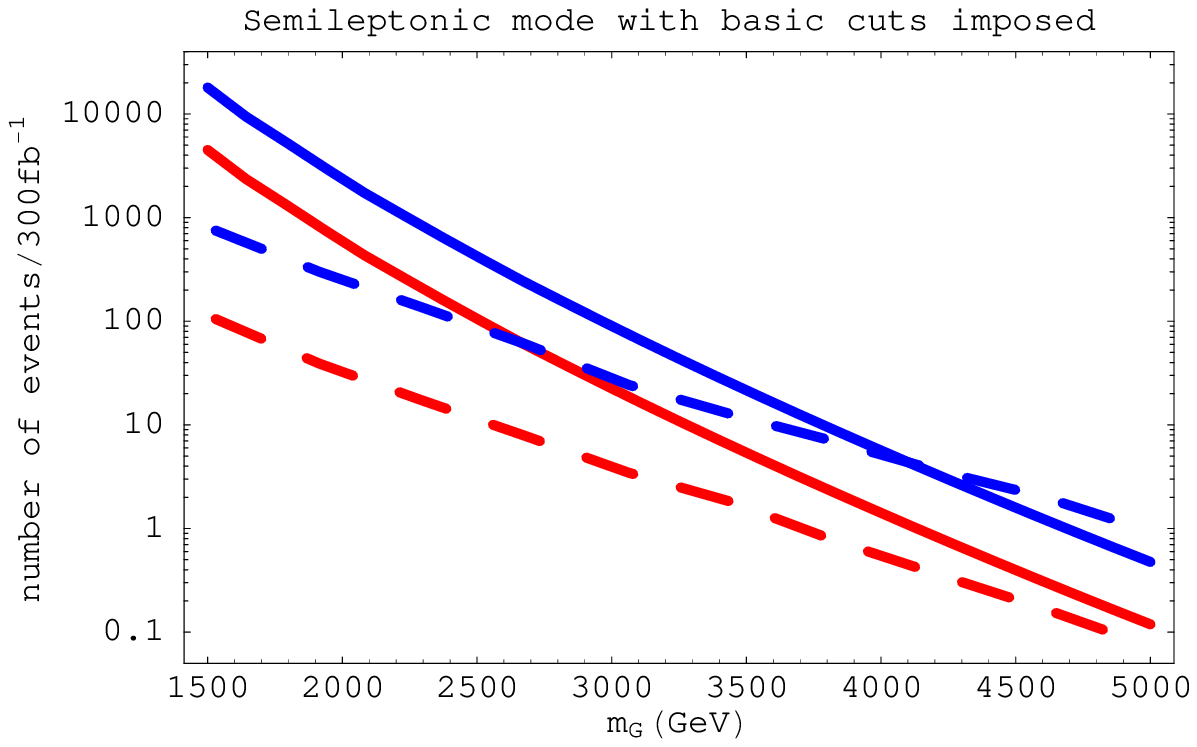}
\centering
\caption{(Color online) (a) The total signal (solid) and SM background (dashed) cross-section (integrated in $m_G \pm \Gamma_G$ window) for $pp \to W(l\nu) W(jj) $ after $|\eta_W|<1$ cuts were applied for c=1 (red) and c=2 (blue) values, (b) Corresponding number of events for 300 fb$^{-1}$.}
\label{WW}
\end{figure}

With efficient hadronic W mass reconstruction, we have another case when most of the hadronic QCD background can be separated; and we are left with WW as the only irreducible background. Fig.\ref{WW} shows results for this situation with $c\equiv k/M_{Pl}=$1 and 2 (see Ref.\cite{Agashe:2007zd} for the discussion of the range of c). Notice that for Fig.\ref{WW}, we integrated over ($m_G \pm \Gamma_G$) WW invariant mass window compared with ($m_G \pm \Gamma_G/2$) window for Fig.\ref{WW1}.

We see that in the c=1(2) case, gravitons up to 3.5 TeV (4 TeV) mass might have enough events to be observed with good statistical significance;
see
Table~\ref{table3}. 
Dependence of the SM WW background on the $c$ value follows from the fact that $m_G \pm \Gamma_G$ integration region is not constant since $\Gamma_G\sim c^2$ .

\begin{table}
\caption{Semileptonic mode signal cross-sections [in fb] and S/B ratios along with W + 1 jet and WW SM backgrounds. Signal 1 and the 
corresponding W + 1 jet background results were obtained 
after cuts in Eqs.\ref{slcuts},\ref{lepen} were imposed 
and $m_G \pm \Gamma_G/2$ integration region was chosen. Signal 2 and 
corresponding WW background results were obtained after $|\eta_W|<$1 
cut and integrated in $m_G \pm \Gamma_G$ window. When the 
number of events is low ($\lsim$ 10),   
Poisson statistics confidence level is considered a more
appropriate statistical 
description and is consequently used.}    
\label{table3}
\begin{center}
\begin{tabular}{|c|c|c|c|c|}
\hline
2 TeV&Cuts&$\#$ of events/300 fb$^{-1}$&S/B&S/$\sqrt{B}$\\
\hline
Signal 1 [c=1]&1.7&510&1.04&23\\
W + 1 jet background [c=1]&1.64&492&&\\
\hline
Signal 2 [c=1]&2.0&600&13.3&90\\
WW background [c=1]&0.15&45&&\\
Signal 2 [c=2]&7.8&2340&7.8&135\\
WW background [c=2]&1.0&300&&\\

\hline
3.5 TeV&Cuts&$\#$ of events/300 fb$^{-1}$&S/B&S/$\sqrt{B}$\hspace{2mm}(CL)\\
\hline
Signal 1 [c=1]&0.01&3&0.33&1\hspace{2mm}(54$ \%\ $)\\
W + 1 jet background [c=1]&0.03&9&&\\
\hline
Signal 2 [c=1]&0.02&6&2.9&4.1\hspace{2mm}(99$ \%\ $)\\
WW background [c=1]&0.007&2.1&&\\
Signal 2 [c=2]&0.07&21&1.4&5.4\\
WW background [c=2]&0.05&15&&\\

\hline
\end{tabular}
\end{center}
\end{table}

\begin{figure}[htb]
\includegraphics[width=5in]{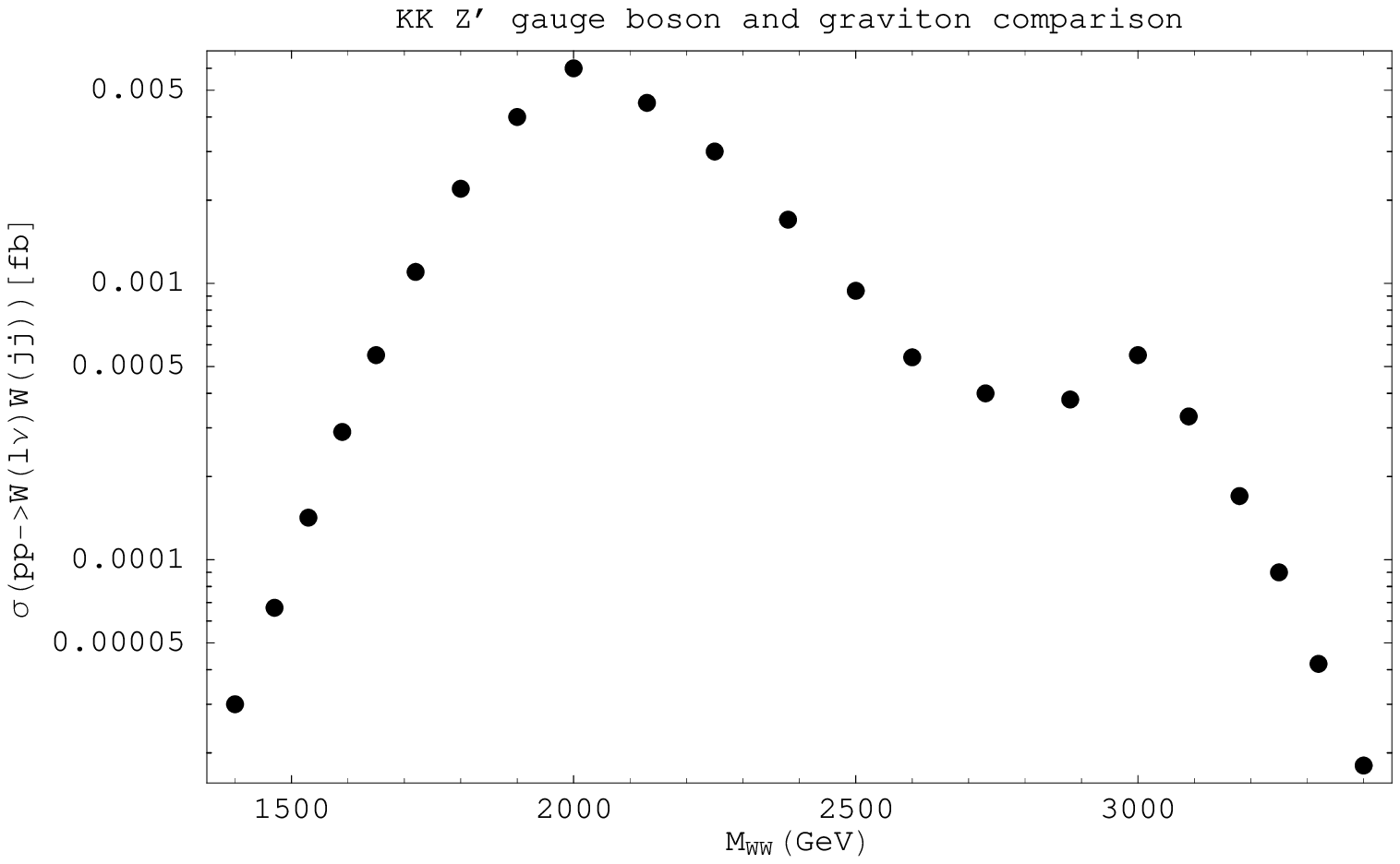}
\centering
\caption{Contributions of the 2 TeV gauge boson and 3 TeV graviton to the $pp \to W(l\nu) W(jj) $ process. Cuts specified in Eq.\ref{slcuts} were applied and $c$=1.}
\label{separ}
\end{figure}

In parallel with leptonic mode, we need to remember that we have a neutral gauge bosons Z$^{\prime}$ produced (through $q\bar{q}$ annihilation or vector boson fusion processes) which might consequently decay to $W_L$ pair \cite{Agashe:2007ki}. Using $m_1^G\approx 1.5 m_1^{Z^{\prime}}$, in Fig.\ref{separ} we show that for 2 TeV $Z^{\prime}$ and corresponding $\sim$ 3 TeV graviton, the signals are well separated as a function of reconstructed WW invariant mass. Thus, by putting a $M_{WW}>3$ TeV cut, the Z$^{\prime}$ signal will become negligible and enhancement in total cross-section is due to graviton only (we obtained graviton cross-section to be 0.04 fb after $M_{WW}>3$ TeV cut). Now, 2 TeV $Z^{\prime}$ can be discovered with 5$\sigma$ statistical significance for an integrated 
luminosity of 100 fb$^{-1}$ in purely leptonic 
channel as was shown in \cite{Agashe:2007ki}, and there the 3 TeV RS graviton 
contribution will be negligible. Thus, assuming that W + 1 jet background will be manageable (for example, by means discussed above) and Z$^{\prime}$ mass is estimated from some other mode (from purely leptonic one we considered above, for example) we may expect to confirm existence of RS graviton in semileptonic channel. Clearly, even if the above relation between masses of these lightest KK modes will not turn out to be true or some other resonance(s) will appear in this channel, WW mass spectrum measurement should 
still provide an important additional information.   

\section{Discussion}

We saw in previous sections that (semi) leptonic modes from $W_L$ pair decay have a potential to discover RS graviton up to about (3.5 TeV) 3 TeV of mass.  To increase statistics, we might expect to use $\tau$ leptons which will give us combinatorial factor of 3  and 3/2 for leptonic and semileptonic modes respectively from 
additional 
decay channels; therefore, the inclusion of the $\tau$'s can help 
appreciably. The reason 
for optimism on the issue of the detection of the $\tau$'s is 
that $\sim$500 GeV energy $\tau$'s have a decay length 
of $l=\gamma \tau c\approx 20$ mm and, thus, might leave visible tracks in the detector \cite{Bengtsson:1985ym}. 
For $m_G \gsim $ 3.5 TeV, higher luminosities are required which will scale our results accordingly \cite{Davoudiasl:2007wf}. Similarly, upgrades of the center of mass energy at LHC \cite{Bruning:2002yh} can extend the reach in KK mass.

So far, the study 
of the RS gravitons was based either on the total cross-section or reconstructed graviton mass measurements. We might try to exploit unique spin-2 nature of the graviton, which might be challenging in our channels. For example, one might be tempted to use lepton pseudorapidity which, due to the high boost of the decaying W's, will be $\sim$sin$^2\theta$ behavior of the basic 2$\to$2 underlying scattering process. But high-energy $W_L^+ W_L^-$ production in the SM and RS Z$^\prime$ decaying to two $W_L$ also have this behavior and, thus, will be indistinguishable in shape from our signal.

Also, we might use information on lepton energy to establish that W's from our graviton decay are longitudinally polarized. This analysis is most promising in semileptonic mode because, as discussed in section  IV B, in this mode WW mass can be reconstructed. Leptons from $W_L$ decay will be preferentially emitted in the direction of the spin axis which is perpendicular to the direction of W motion. Thus, lepton and neutrino will tend to have the same energy in the lab frame, compared to decay of transversely polarized W's where they are emitted in the direction of the W motion and, thus, one of the W decay products will carry most of the energy. Now, suppose we have a negatively charged W decay. To confirm that W's from our graviton decay are longitudinally polarized, we calculated the average lepton energy in the lab frame from the decay of polarized W bosons and summarized our results in Table \ref{table4}. We notice that the average for the longitudinally and transversely polarized W's (which is the average of left-handed and right-handed polarizations) is the same and equal to $\sqrt{s}/4$. To distinguish between longitudinal and transverse polarizations, we divide signal events into two groups: events in the first group will have charged lepton energy bigger than the neutrino's energy; and in the second group, the neutrino's energy will be bigger. The fact that lepton's energy is bigger (smaller) implies that lepton's 3-momentum in the W rest frame is parallel (antiparallel) to W's 3-momentum in the lab frame. We calculated the average lepton energy in the lab frame from the decay of polarized W bosons for the events in each group and presented our results in Table \ref{table4} as well. We see that this analysis could be used to confirm that W's from our graviton decay are longitudinally polarized as, presumably, average lepton energies  will match the $(8+ 3\beta)\sqrt{s}/32$ and $(8- 3\beta)\sqrt{s}/32$ values for the signal events in the first and second group respectively. For a positively charged W decay, the results for the left-handed and right-handed rows in Table \ref{table4} need to be switched. 

\begin{table}
\caption{Average lepton energies in the lab frame from the decay of polarized W$^-$ bosons. For a W$^+$ decay, the results for the left-handed and right-handed rows need to be switched.}
\label{table4}
\begin{center}
\begin{tabular}{|c|c|c|c|}
\hline
W polarization&Average& Average for group 1& Average for group 2\\
\hline
Longitudinal&$\sqrt{s}/4$&$(8+ 3\beta)\sqrt{s}/32$&$(8- 3\beta)\sqrt{s}/32$\\
\hline
Left-handed&$(2+\beta)\sqrt{s}/8$&$(28+ 17\beta)\sqrt{s}/112$&$(28- 17\beta)\sqrt{s}/112$\\
\hline
Right-handed&$(2-\beta)\sqrt{s}/8$&$(28+ 17\beta)\sqrt{s}/112$&$(28- 17\beta)\sqrt{s}/112$\\
\hline
\end{tabular}
\end{center}
\end{table}

\section{Conclusion}

In this work, we have extended earlier studies of the discovery potential of warped gravitons at the LHC which concentrated on the gravitons decaying into the ``gold-plated'' ZZ channel and into the $t\bar{t}$ pair. We have considered resonant production of the first RS KK graviton mode via gluon-fusion process followed by its subsequent decay to $W_L W_L$ pair. We focused on leptonic and semileptonic final states and found that with 300$fb^{-1}$ of data, 
LHC 
may discover 
first RS KK graviton with masses below $\sim$ 3 TeV and 3.5 TeV in these modes respectively. We also incorporated potential KK $Z^{\prime}$ signal in both modes and analyzed its combined effect with RS graviton. Taking the RS prediction for the 
lightest KK masses, 
$m_1^G\approx 1.5 m_1^{Z^{\prime}}$, we showed 
that these signals are well separated in reconstructed WW invariant mass in the semileptonic mode. For the purely leptonic $e\mu$ mode, where resonance mass reconstruction is problematic, the
above mass relationship hints to the domination of the $Z^{\prime}$ events as corresponding graviton mass will be higher. Nevertheless, we demonstrated that even in that mode, appropriate choice of cuts in dilepton invariant mass may be able to distinguish these contributions as well. 

\begin{acknowledgments}

We thank H. Davoudiasl for a careful reading of the manuscript
and to him and to J.Cochran for useful discussions. O.A. also would like to thank BNL Physics Department for hospitality during part of this project. Work of O.A. and D.A. are supported in part by DOE under contract number DE-FG02-01ER41155. A.S. is supported in part by the DOE grant DE-AC02-98CH10886 (BNL).

\end{acknowledgments}


\begin{thebibliography}{999}
\bibitem{tau}
In all the numerical estimates presented in this paper, we will only
include $W \to e \nu_e$ and $W \to \mu \nu_\mu$. However, towards the
end in Section VI we will emphasize that the detection
of the $\tau$'s from these decays may indeed be feasible.
\end{thebibliography}
\end{document}